# APPLICATIONS OF GEOMETRIC ALGORITHMS TO REDUCE INTERFERENCE IN WIRELESS MESH NETWORK

Hung-Chin Jang

Department of Computer Science, National Chengchi University, Taiwan
jang@cs.nccu.edu.tw

*ABSTRACT*

*In wireless mesh networks such as WLAN (IEEE 802.11s) or WMAN (IEEE 802.11), each node should help to relay packets of neighboring nodes toward gateway using multi-hop routing mechanisms. Wireless mesh networks usually intensively deploy mesh nodes to deal with the problem of dead spot communication. However, the higher density of nodes deployed, the higher radio interference occurred. This causes significant degradation of system performance. In this paper, we first convert network problems into geometry problems in graph theory, and then solve the interference problem by geometric algorithms. We first define line intersection in a graph to reflect radio interference problem in a wireless mesh network. We then use plan sweep algorithm to find intersection lines, if any; employ Voronoi diagram algorithm to delimit the regions among nodes; use Delaunay Triangulation algorithm to reconstruct the graph in order to minimize the interference among nodes. Finally, we use standard deviation to prune off those longer links (higher interference links) to have a further enhancement. The proposed hybrid solution is proved to be able to significantly reduce interference in a wireless mesh network in O(n log n) time complexity.*

## 1. INTRODUCTION

In recent years, wireless mesh networks (WMN) have become more popular than ever. Many technologies, like IEEE 802.11s, IEEE 802.15 (multi-hop mode), IEEE 802.16 (mesh mode) and wireless sensor network, etc., are of this kind. A multi-hop WMN can not only extend coverage but also save both cabling cost and human resource. WMN is essentially robust in fault tolerance. Even if some of the mesh nodes are incapable, there exist many other alternative nodes to help relay. It is demanding a full mesh to have the best fault tolerance. However, this condensed mesh nodes deployment may cause significant packet collisions. Gupta [4] found that given *n* identical nodes, each node is within each other' communication range and is in the same collision domain. If each node is capable of transmitting *W* bits per second (bps) then the throughput will be $\Theta\left(\frac{W}{\sqrt{n}}\right)$ bps. If *n* nodes are randomly deployed, the average throughput of each node is $\Theta\left(\frac{1}{\sqrt{n \log n}}\right)$, and the optimal throughput of each node will be $\Theta\left(\frac{1}{\sqrt{n}}\right)$. Burkhart [3] gave a concise and intuitive definition of interference. Kodialam [8] investigated the throughput capacity of wireless networks between given source destination pairs using a simple interference model. Kumar [9] considered the same problem for various interference models. However, they both do not take channel allocation into account as they consider a single-channel network. Mesh routers with multiple-radios can transmit and receive simultaneously or can transmit on multiple channels simultaneously seems to be a feasible solution. Gupta [4] pointed out that one of the major problems facing wireless networks is the capacity reduction due to interference among multiple simultaneous transmissions. Alicherry [1] believed that the interference cannot be completely eliminated due to the limited number of channels available and suggested that careful channel assignment must be done to mitigate the effects of interference. Kyasanur [10] proposed algorithms for channel assignment and routing in multi-channel multiple Network Interface Card (NIC) MANETs. Subramanian [18] assigned channels to communication links to minimize the overall interference in a multiradio wireless mesh network. Tang [19] used topology control and QoS routing in multi-channel wireless mesh networks to deal with interference problem. Jain [7] used conflict graph to model wireless interference and detected the group of links that might have significant interference. Jain further proved that the problem of finding optimal throughput is NP-hard. Padhye [13] and Reis [16] proposed to use measurement-based techniques to derive conflict graphs. Ramachandran [14] proposed a purely measurement-based approach for channel assignment to radios. Moaveninejad [12] extended the results from [3] and proposed algorithms for constructing a network topology in wireless ad hoc networks such that the maximum (or average) link (or node) interference of the topology is either minimized or approximately minimized. Hsiao [5] identified the links intersection problem of nodes using directional antennas in a wireless network. Hsiao showed that if a line intersection occurs and the four ends (nodes) of the two intersected lines are communicating at the same time then all the four nodes will interfere with each other. Burkhart [3] showed that low degree of nodes does not guarantee low interference in a wireless network. Since the transmission range of a node is determined by the longest path, one node with low degree may have large transmission range. Once it transmits data, it may interfere the transmissions of the others. Some proposals originated in computational geometry, such as the Delaunay Triangulation [6], the minimum spanning tree [15] and the Gabriel Graph [17]. Hu [6] applied Delaunay Triangulation (DT) algorithm to solve network problem. Delaunay Triangulation algorithm is able to maximize all the angles in a geometric graph and thus shorten the lengths of links. Delaunay Triangulation algorithm is able to reduce the total length of links while keep the same total degrees of nodes. Mapping to the network problem, reducing lengths of links implies reducing transmission ranges among nodes. This may help to mitigate the interference problem. Figure 1(a) shows an original network topology modeled by a graph. Figure 1(b) is the triangulated network topology applying Delaunay Triangulation algorithm. It shows that many links in Figure 1(b) are shorter than those in Figure 1(a). Meguerdichian [11] applied Voronoi diagram and Delaunay Triangulation algorithm to wireless sensor networks to optimize sensors deployment.

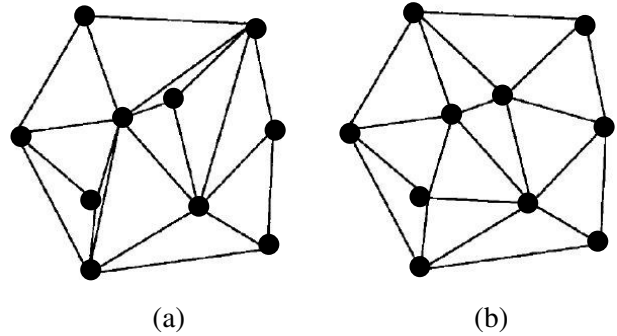

Figure 1. (a) Original topology (b) Triangulated topology.

In this paper, we extend links intersection problem of directional antennas to that of omni-directional antennas, and focus on enhancing network performance by minimizing the interference among nodes. We convert network problem into geometry problem by defining line intersection in a graph to reflect radio interference problem. We then use plan sweep algorithm to find if there exists any intersection lines. We further employ Voronoi diagram algorithm and Delaunay Triangulation algorithm to reconstruct the graph in order to minimize the interference among nodes. Finally, we use standard deviation to prune off those longer links (higher interference links) to have a further enhancement of network performance. The paper organization is as follows. Section 2 is the framework of the proposed solution. Section 3 shows the simulations and results using NS2 (Network Simulator 2). Section 4 offers brief concluding remarks.

## 2. FRAMEWORK

The operation flow of the proposed interference reduction framework is as follows (Figure 2). In the beginning, nodes deployment and positioning is used to deploy mesh nodes in a target area. Positioning can be done by either of GPS, AOA (Angle of Arrival), TDOA (Time Difference of Arrival) or RSSI (Received Signal Strength Indicator), etc. Locations broadcast is used to broadcast nodes coordinates to their neighbors. Hence, each node will be able to know the coordinates of those nodes within its transmission range. Network problems are converted to geometry problems and then these problems are solved by geometric algorithms. Those algorithms used by the framework are plane sweep algorithm, Voronoi diagram algorithm and Delaunay Triangulation algorithm. Plan sweep algorithm is used to check line intersections. Voronoi diagram algorithm is used to search neighboring nodes. Delaunay Triangulation algorithm is used to reduce the lengths of links among nodes (or to reduce the interference among nodes). Finally, we use standard deviation to prune off those longer links (or higher interference links) to have a further enhancement of network performance. Transmission power of sensors will thus be adjusted according to the results of these geometric algorithms. This completes an iteration of interference reduction process. Since the topology is apt to changed, nodes insertion and deletion is required to keep the topology up to date.

## 2.1 Line Intersection Problem

Line intersection in geometry can be mapped to the interference problem in a real network when two intersected pairs of communication nodes are transmitting data at the same time. If the communication between one of these two pairs can be redirected to the other nodes, it is possible not only to avoid collisions but also reduce both the transmission radii of communication nodes and the number of interfered nodes. Therefore, it is possible to largely reduce communication interference by removing some intersected lines in graph. Figure 3 shows that there are nodes A, B, C and D in a WMN and each node is equipped with an omni-directional antenna. We assume that node A is communicating with node B and node C is communicating with node D at the same time. Without loss of generality, segment AB may be longer than or equal to segment CD. In the first case, the transmission range of node A will cover node D and C, in the second case, node A, B, C or D are able to cover each other. In either case, interference can be reduced by removing one of these two intersected segments. Figure 4(a) shows that if any node of a fully mesh WMN wants to send messages to another node then all the other nodes must stay in listening mode. This may significantly reduce system throughput. If Figure 4(a) is transformed into Figure 4(b) by removing intersected lines then nodes F and B are able to send messages to node E and C, respectively, at the same time.

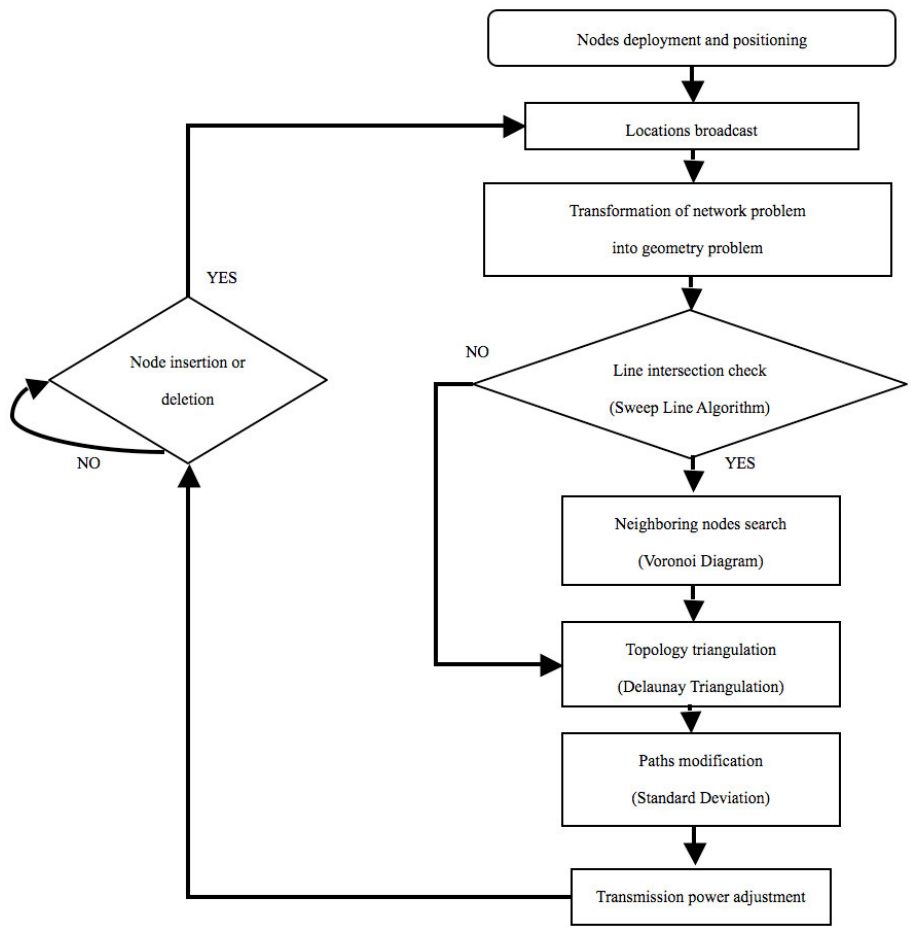

Figure 2. Interference reduction framework.

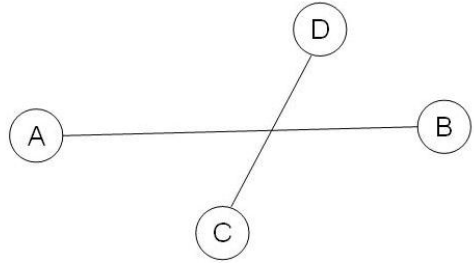

Figure 3. Example of line intersection.

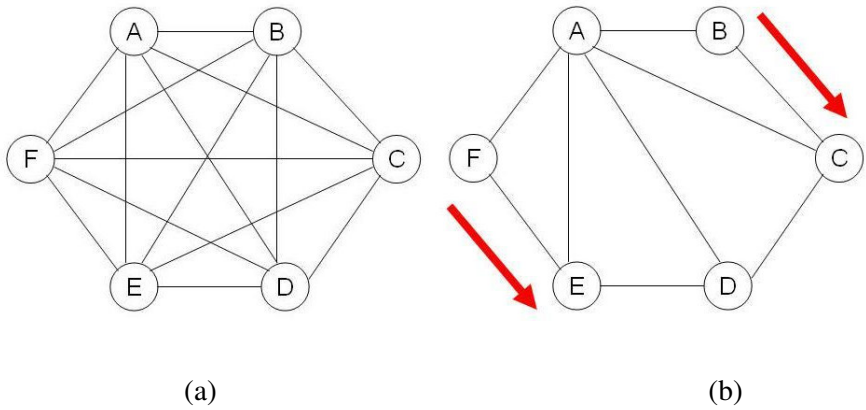

(a)                               (b)

Figure 4. Six nodes of fully mesh and triangulated mesh.

## 2.2 Plane Sweep Algorithm

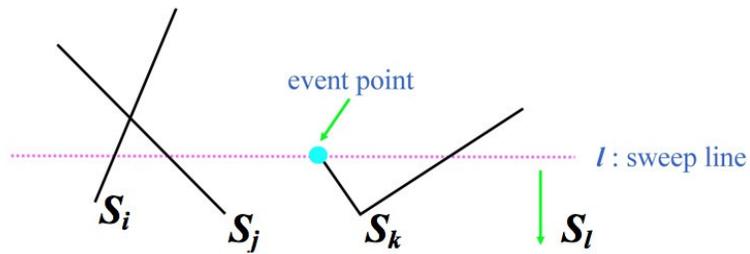

Figure 5. Sweep line and event point.

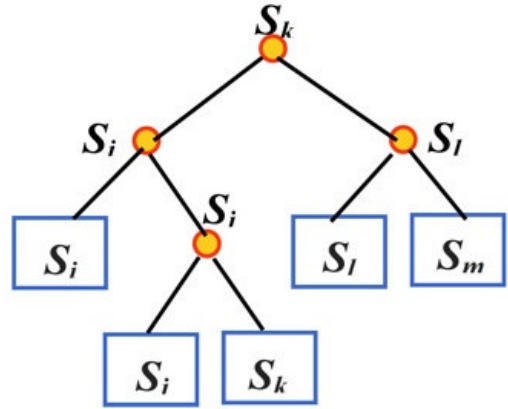

Figure 6. Data structure of an event queue.

Plane sweep algorithm is used to check line intersections in a geometric graph. Plane sweep algorithm is realized as follows. All the lines are numbered in advance. Scan all the lines in the graph from top to bottom (Figure 5). Once there is an endpoint of a line found, say an "event point", we add this point to an "event queue", a balanced binary tree (Figure 6). The event queue is used to maintain the event points of line intersection check. An event point won't be removed from an event queue until the line of that event point passes line scan. Plane sweep algorithm has time complexity, $O((n+I) \log n)$, where $n$ is the number of event points and $I$ is the number of intersection points. As a comparison, the time complexity of that of using brute force algorithm is $O(n^2)$ instead.

## 2.3 Voronoi Diagram Algorithm

In a Voronoi diagram, it holds the property that the nearest site of any point $x$ in a sub-area $V(P_i)$ must be $P_i$ (site). Voronoi diagram algorithm is used here to find the dividing lines and adjacent nodes. The time complexity of Voronoi diagram is $O(n \log n)$, where $n$ is the number of nodes. The definition of Voronoi diagram is as follows. Let $P = \{P_1, P_2,...,P_n\}$, $n \geq 2$, where $P$ is a set of nodes in an area, and $P_1$, $P_2$, ..., $P_n$ are sites. Let $V(P)=\{V(P_1),V(P_2),......,V(P_n)\}$, where $V(P_i)=\{x: P_i\text{-}x \leq P_j\text{-}x, j \neq i\}$. $V(P)$ is called a Voronoi diagram.

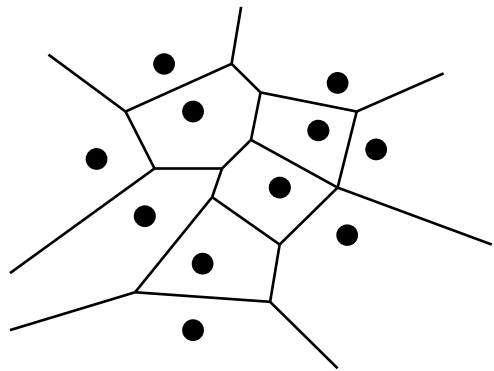

Figure 7. Nodes partitioned by Voronoi diagram.

## 2.4 Delaunay Triangulation Algorithm

Delaunay Triangulation algorithm is used to reduce the lengths of links among nodes. This algorithm will be applied to our problem in two cases. First is in the case of original topology transformation. Second is in the case of topology reconstruction after nodes addition and deletion. Delaunay Triangulation algorithm uses Legal Triangulation [2] algorithm to optimize polygons. Figure 8 shows two cases of triangulated polygons. If the sum of angles $\alpha_2$ and $\alpha_6$ is greater than that of $\alpha_1'$ and $\alpha_6'$ then edge $P_iP_j$ is said to be an "illegal edge" and the flipped edge $P_lP_k$ is said to be a "legal edge". Legal Triangulation algorithm flips an "illegal edge" into a "legal edge". The time complexity of Delaunay Triangulation diagram is $O(n \log n)$, where $n$ is the number of nodes.

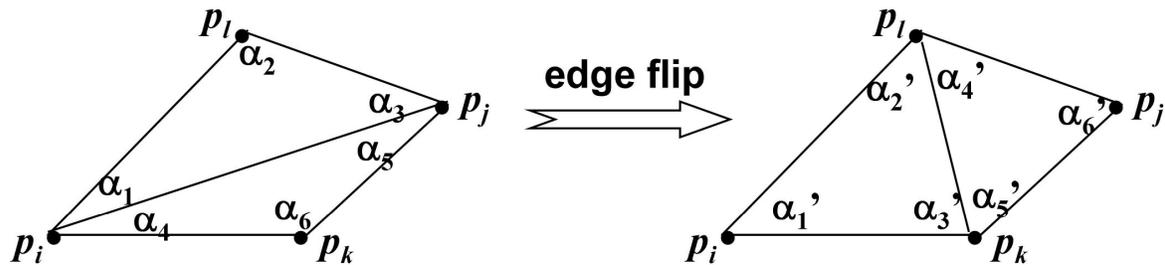

Figure 8. Example of Delaunay Triangulation.

As nodes deployed in different locations, each node may either have "direct interference", "indirect interference", or "no interference" on other nodes. Direct interference means that once a node wants to transmit data, all those nodes within its transmission range should stay in listening mode to avoid collision. Indirect interference means that once a node wants to transmit data to its neighboring nodes, those nodes which would like to communicate with these neighboring nodes must stay in listening mode. No interference means that once a node wants to transmit data, there is no worry about collision to those neighboring nodes. Given a full mesh network of six nodes in Figure 9, three possible triangulated graphs by Delaunay Triangulation algorithm are shown in Figure 10. Comparing Table 1 with Tables 2 to 4, we find the variations of interference effects on neighboring nodes given distinct network topologies. Triangulated graphs can really mitigate the interference problem to a certain degree.

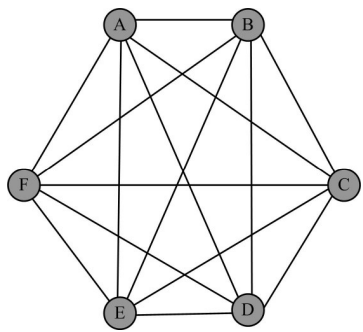

Figure 9. Full mesh network of six nodes.

Table 1. Interference effects on nodes in Figure 9.

|  | Node A (similar to B, C, D, E, F) |
|---|---|
| Direct interference | B, C, D, E, F |
| Indirect interference | Nil |
| No interference | Nil |

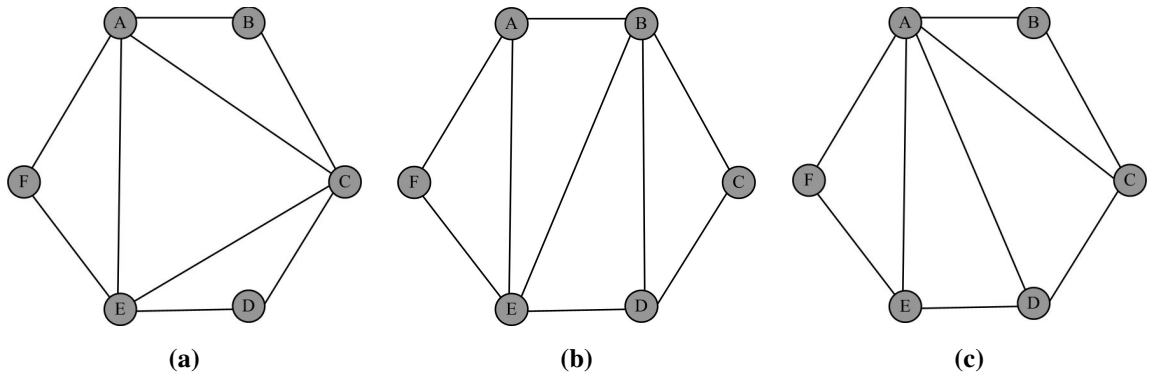

(a)          (b)          (c)

Figure 10. Triangulated graph by DT algorithm.

Table 2. Interference effects on nodes in Figure 10(a).

|  | Node A (similar to C, E) | Node B (similar to D, F) |
|---|---|---|
| Direct interference | B, C, E, F | A, C |
| Indirect interference | D | D, E, F |
| No interference | Nil | Nil |

Table 3. Interference effects on nodes in Figure 10(b).

|  | Node A (similar to D) | Node B (similar to E) | Node C (similar to F) |
|---|---|---|---|
| Direct interference | B, E, F | A, C, D, E | B, D |
| Indirect interference | C, D | F | A, E |
| No interference | Nil | Nil | F |

Table 4. Interference effects on nodes in Figure 10(c).

|  | **Node A** | **Node B** (similar to F) | **Node C** (similar to E) | **Node D** |
| --- | --- | --- | --- | --- |
| Directly affect | B, C, D, E, F | A, C | A, B, D | A, C, E |
| Indirectly affect | Nil | D, E, F | E, F | B, F |
| No affect | Nil | Nil | Nil | Nil |

## 2.5 Path Modification

To keep the property of connectivity, path modification should be based on clustering. A group of close related nodes are grouped into one cluster. There exists at least one path between two clusters. In a geometric graph, the degree of a node equals to the number of paths (links). The longest path of a node is thought to be the transmission range. Figure 11 shows that path AB is the longest link for both node A and node B. Hence, the transmission range of node A and node B is the length of AB. When node A or node B starts to transmit data, all the other nodes should stay in listening mode since they all are within the transmission range of radius AB. To reduce such interference, we apply standard deviation to prune off those relatively longer links.

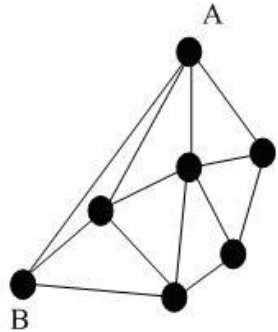

Figure 11. Path AB is the longest link of both node A and node B.

In deciding which links to remove, we classify the standard deviation ranges at both endpoints of a link into four levels as shown in Table 5. We also associate probability with each level of range. We set Level 0 and Level 3 have the highest and the least probabilities of occurrences, respectively. Since both endpoints of one link may not fall on the same level, the priority of link pruning depends on the joint level of standard deviation ranges at both endpoints. The priority is set according to Table 6. In addition, if two links have the same link pruning priority, the link covers more nodes will be removed first. For example, if a node is of degree 9 and each of its links has different lengths and different number of covered nodes as shown in Table 7. The priorities of links to be removed are numbered as follows: 2, 4, 8, 5, 7, 6, 3, 9, 1.

Table 5. Four levels of standard deviation range.

| Standard deviation range | Level | Probability |
|---|---|---|
| $< \mu$ | Level 0 | 50% |
| $\mu \sim \mu+\sigma$ | Level 1 | 34.2% |
| $\mu+\sigma \sim \mu+2\sigma$ | Level 2 | 13.6% |
| $> \mu+2\sigma$ | Level 3 | 2.2% |

Table 6. Priority of link pruning

| Standard deviation range (endpoint 1) | Standard deviation range (endpoint 2) | Priority |
|---|---|---|
| Level 3 | Level 3 | 1 |
| Level 3 | Other (Level 0, 1, 2) | 2 |
| Other (Level 0, 1, 2) | Level 3 | 2 |
| Level 2 | Level 2 | 3 |
| Level 2 | Other (Level 0, 1) | 4 |
| Other (Level 0, 1) | Level 2 | 4 |
| Other (Level 0, 1) | Other (Level 0, 1) | 5 |

Table 7. Link information of a node of degree 9.

| Link no. | Length of link | Level of standard deviation | No. of covered nodes |
|---|---|---|---|
| 1 | 12 | Level 0 ($< \mu$) | 2 |
| 2 | 250 | Level 3 ($> \mu+2\sigma$) | 12 |
| 3 | 22 | Level 0 ($< \mu$) | 4 |
| 4 | 210 | Level 2 ($\mu+\sigma \sim \mu+2\sigma$) | 11 |
| 5 | 36 | Level 0 ($< \mu$) | 7 |
| 6 | 30 | Level 0 ($< \mu$) | 5 |
| 7 | 33 | Level 0 ($< \mu$) | 6 |
| 8 | 120 | Level 1 ($\mu \sim \mu+\sigma$) | 8 |
| 9 | 15 | Level 0 ($< \mu$) | 3 |
| $\mu = 80.9$ | | $\mu = 85.8$ | |

## 2.6 Nodes Insertion and Deletion

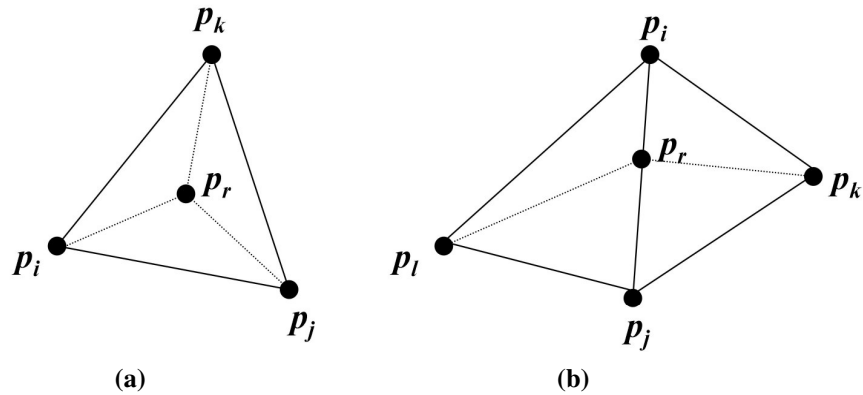

(a)          (b)

Figure 12. Examples of node insertion.

As a node is to be inserted into a network, it has to broadcast its own coordinates to neighboring nodes. The system is then triggered to execute plane sweep algorithm to re-sweep and execute Delaunay Triangulation algorithm to reconstruct network topology. The time complexity of node insertion is $O(n \log n)$. Figure 12 shows two examples of node insertion. Figure 12(a) is the case when the new node $P_r$ falls in the area of a triangle, and Figure 12(b) is the case when $P_r$ lies on the border of two triangles. Figure 13 shows the topology reconstruction after node insertion. Once a node fails or moves out, it triggers node deletion and then topology reconstruction. Node deletion is triggered once the neighboring nodes of a certain node do not receive periodic location update from that node after a timeout.

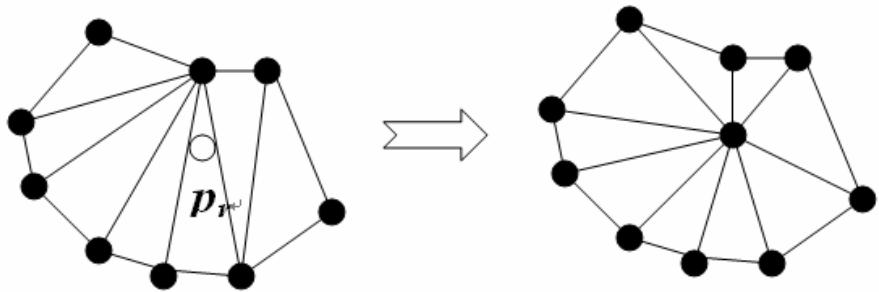

Figure 13. Topology reconstruction.

## 3. SIMULATIONS AND RESULTS

### 3.1 Simulations Environment and Parameters Setting

Simulations are conducted under the following network environment and parameters setting. The pre-process of simulations is as follows: J2SE is used to generate 50, 100, 150, 200 random deployed nodes, respectively. Each node has 300 meters of transmission range. The initial 100 nodes deployment is shown in Figure 14(a). We then use Voronoi diagram algorithm to delimit regions among nodes as shown in Figure 14(b). Figure 14(b) is then triangulated into Figure 14(c) using Delaunay Triangulation algorithm. Figure 14(d) is derived from Figure 14(c) by pruning off longer links based on the result of standard deviation of 100 nodes (Figure 15).

Table 8. Simulations setting.

| Parameters | Description |
|---|---|
| Network environment | 1. simulated outdoor random deployment <br> 2. area: 1000 (m) × 1000 (m) <br> 3. no. of nodes: 50, 75, 100, 125, 150, 175, 200 nodes |
| Node capability | 1. initial transmission range: 300 meters <br> 2. is able to adjust its own transmission power <br> 3. equipped with omni-directional antenna <br> 4. fixed position |
| Development tools | Java 2 Platform Standard Edition (J2SE) programming language is responsible for the following tasks: <br> 1. random deployment <br> 2. Sweep line algorithm <br> 3. Voronoi diagram algorithm <br> 4. Delaunay Triangulation algorithm <br> 5. Standard Deviation rule <br> 6. establishment of random FTP connections |
| Simulator | Network Simulator ver. 2 |

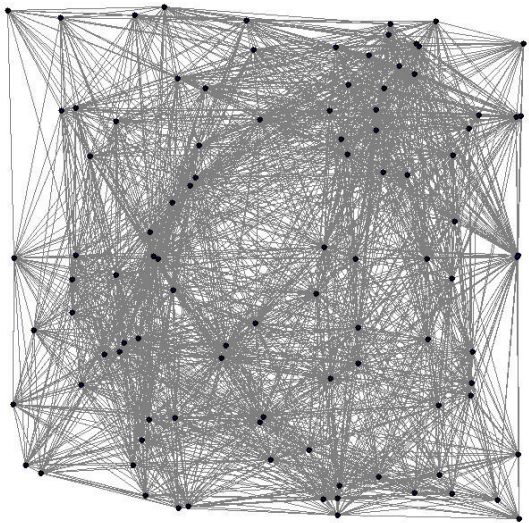
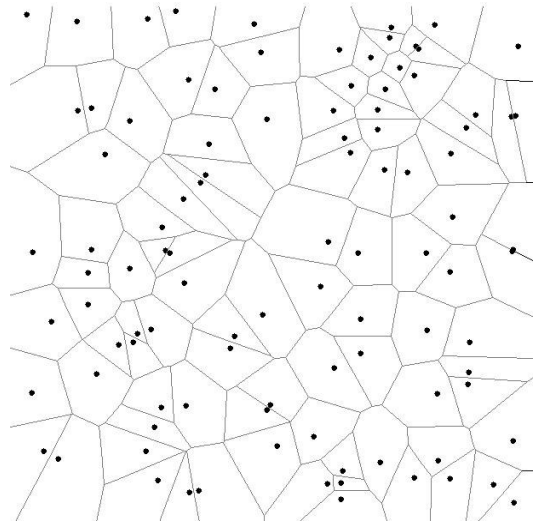

(a) Initial nodes deployment          (b) Regions delimitation

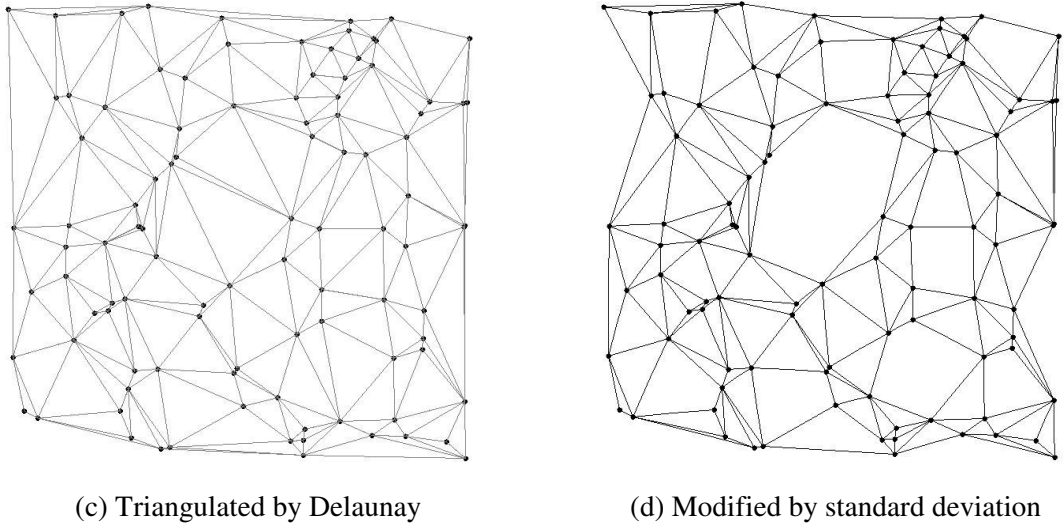

(c) Triangulated by Delaunay        (d) Modified by standard deviation

Figure 14. Network topology with 100 nodes.

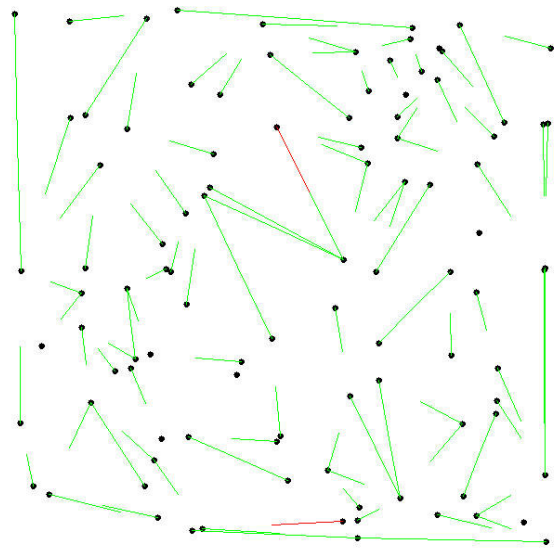

Figure 15. The result of Standard Deviation of 100 nodes.

## 3.2 Performance Comparisons   Node Degree, Transmission Range, Interference Rate

This category of simulations is to compare node degree, transmission range and interference rate of "DT only (Delaunay Triangulation only)" and "DT+SD (Delaunay Triangulation with Standard Deviation enhancement)" methods according to 50, 75, 100, 125, 150, 175, 200 different numbers of mesh nodes. For each different number of mesh node, we set up 10 different network topologies. Figure 16 and 17 show the comparisons of total degree and average degree of nodes of "DT only" and "DT+SD" methods, respectively. We see that the average degree of nodes increases as the number of nodes increases. Besides, "DT+SD" outperforms "DT only" by 8.9% less average degree of nodes on average. Figure 18 shows the comparison of transmission range of "DT only" and "DT+SD" according to different number of nodes. We see that the transmission range decreases as the number of nodes increases. And, "DT+SD" outperforms "DT only" by reducing 13.3% transmission range on average. Figure 19 shows the comparison of average interference rate of "DT only" and "DT+SD" according to different number of nodes. We see that the average interference rate reduces as the number of nodes increases. And, "DT+SD" outperforms "DT only" by 12.3% less interference rates on average.

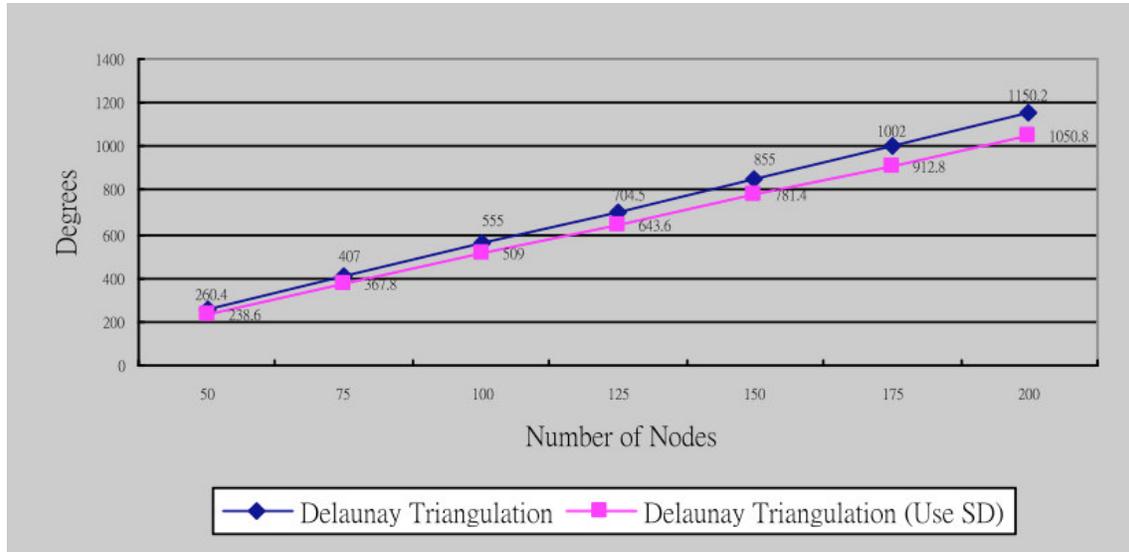

Figure 16. Comparison of total degree of nodes of "DT only" and "DT+SD".

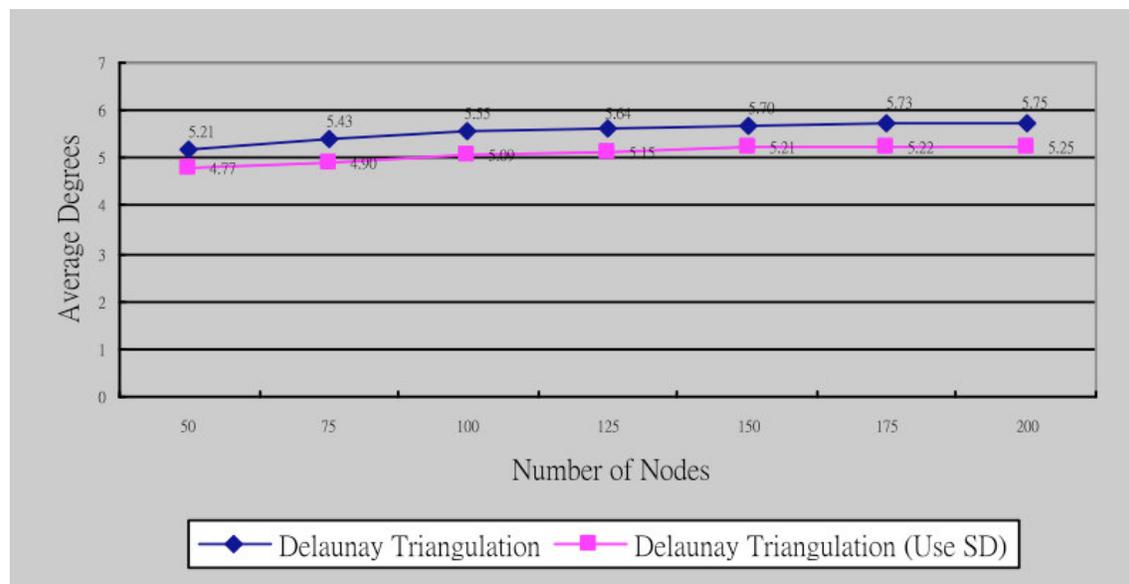

Figure 17. Comparison of average degree of nodes of "DT only" and "DT+SD".

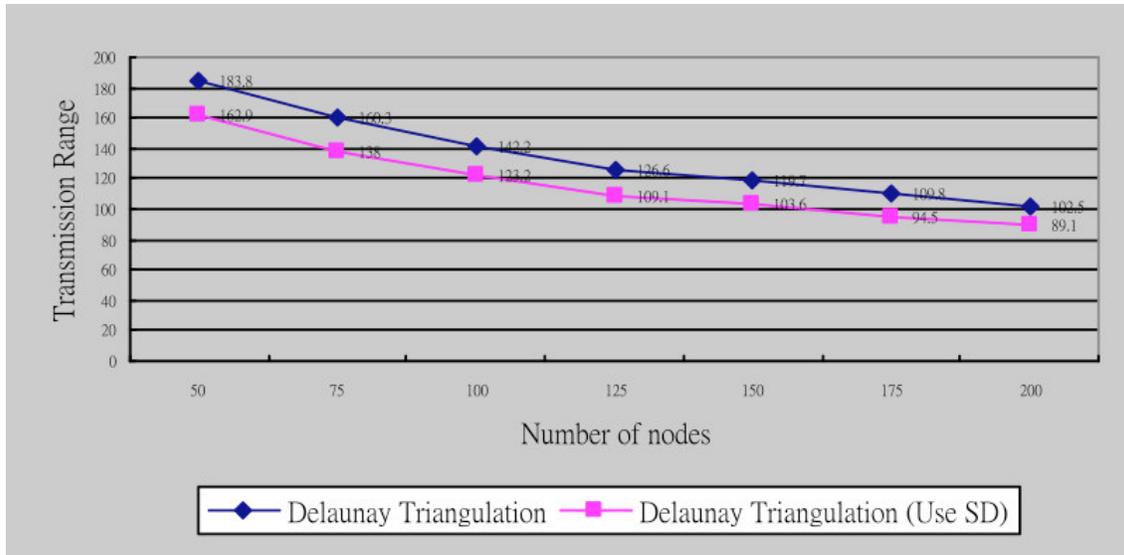

Figure 18. Comparison of transmission range of "DT only" and "DT+SD".

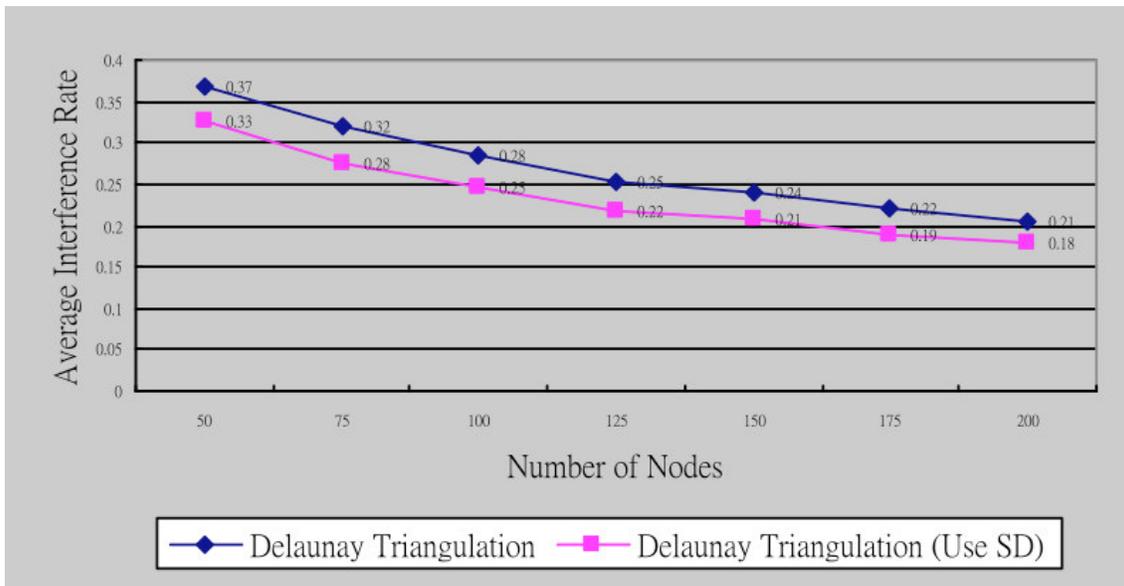

Figure 19. Comparison of average interference rate of "DT only" and "DT+SD".

### 3.3 Performance Comparisons throughput, packet loss rate, delay time

This category of simulations is to compare throughput, packet loss rate and delay time of "Original", "DT only" and "DT+SD" methods according to the following scenarios: "50 nodes with 30 random incoming ftp connections", "100 nodes with 50 random incoming ftp connections", "150 nodes with 100 random incoming ftp connections", and "200 nodes with 150 random incoming ftp connections", respectively. Figures 20-31 show the comparisons of throughput (bps) (Figures 20-23), packet loss rate (Figures 24-27) and delay time (Figures 28-31) of "Original", "DT only", and "DT+SD" according to different number of topologies based on different scenarios. All these results show that "DT+SD" outperforms both "Original" and "DT only" on average.

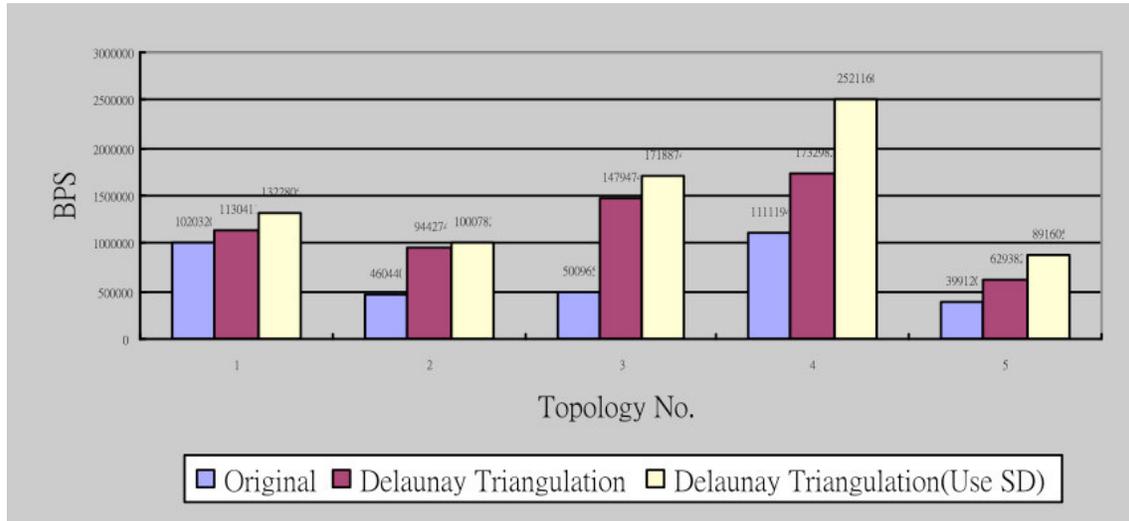

Figure 20. Comparison of throughput ("50 nodes with 30 random incoming ftp connections").

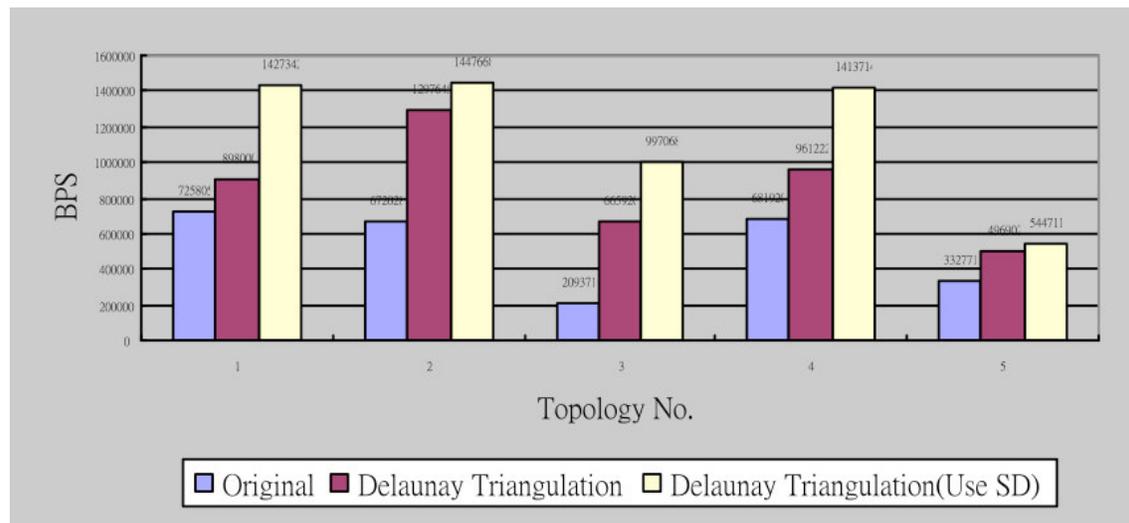

Figure 21. Comparison of throughput ("100 nodes with 50 random incoming ftp connections").

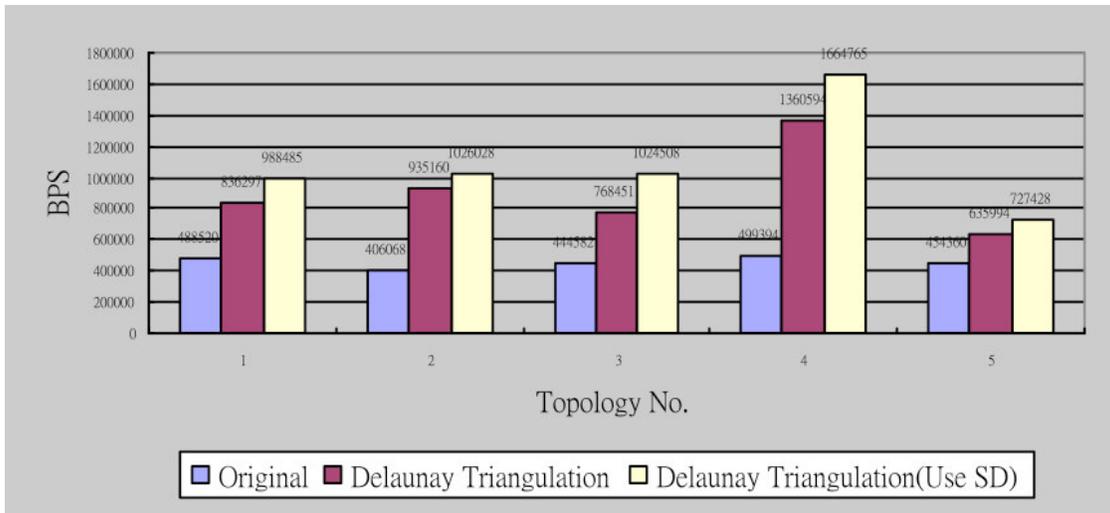

Figure 22. Comparison of throughput ("150 nodes with 100 random incoming ftp connections").

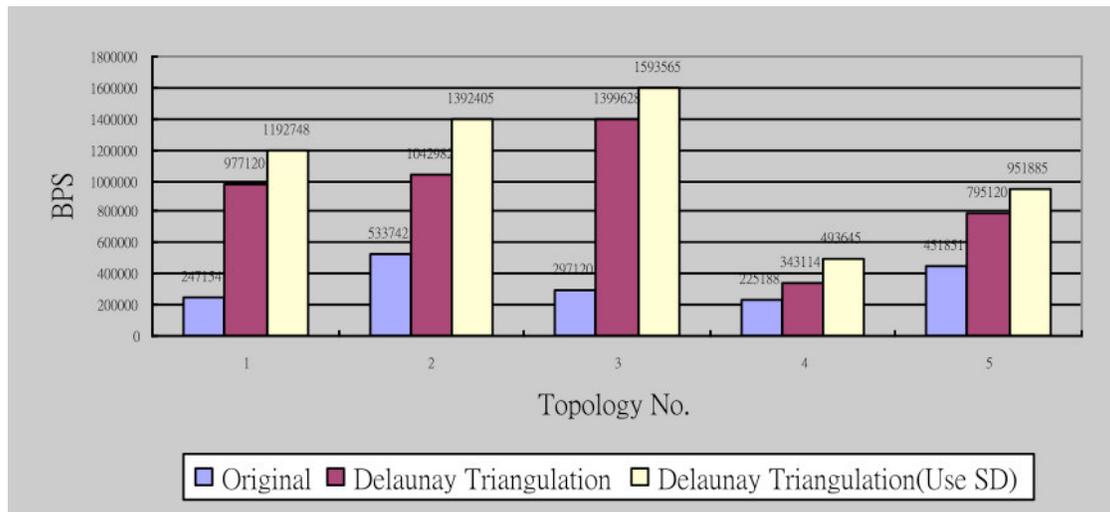

Figure 23. Comparison of throughput ("200 nodes with 150 random incoming ftp connections").

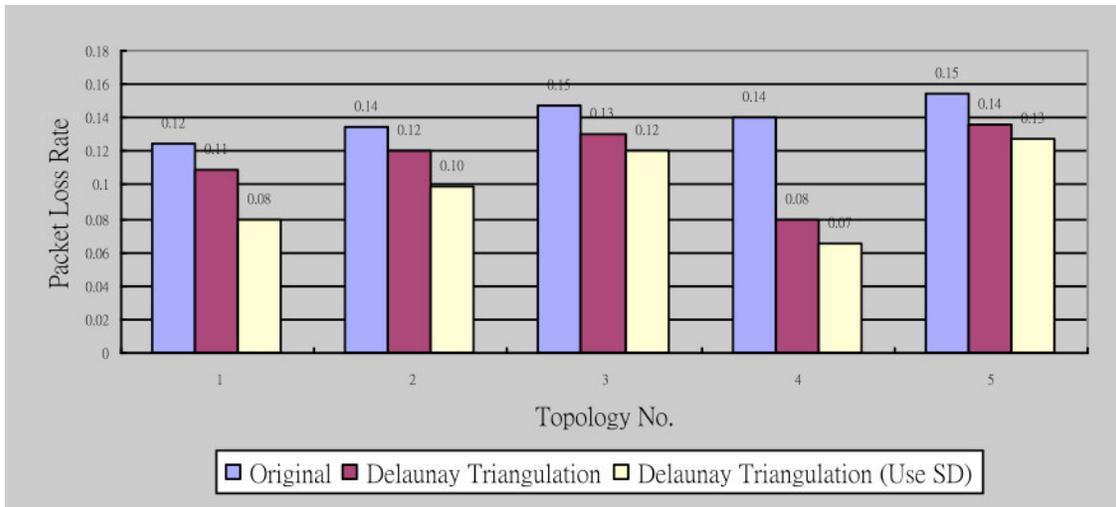

Figure 24. Comparison of packet loss rate

("50 nodes with 30 random incoming ftp connections").

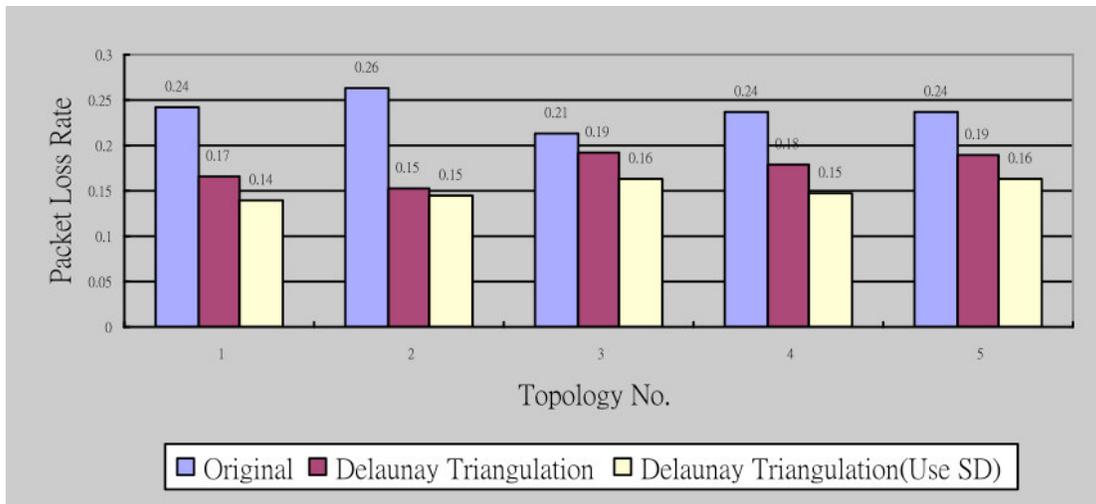

Figure 25. Comparison of packet loss rate

("100 nodes with 50 random incoming ftp connections").

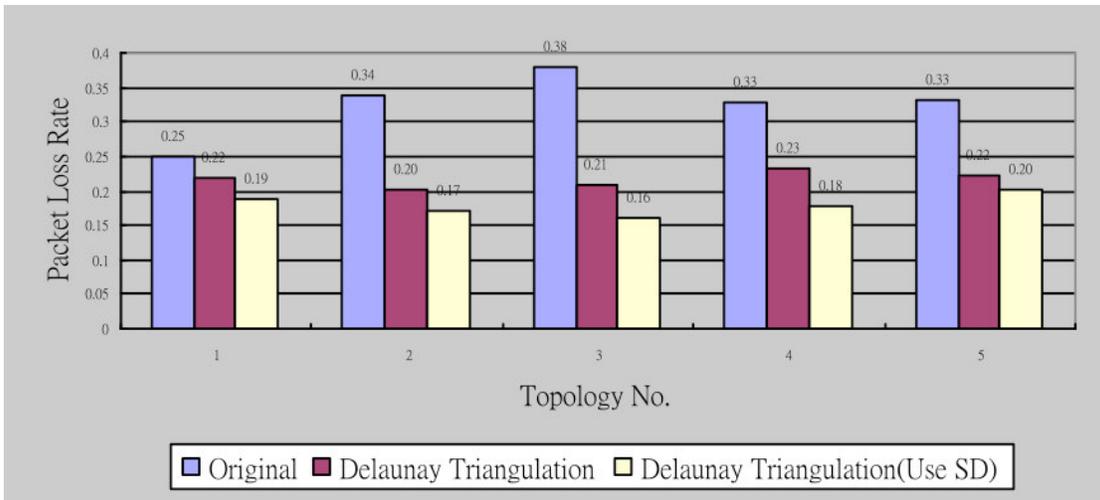

Figure 26. Comparison of packet loss rate

("150 nodes with 100 random incoming ftp connections").

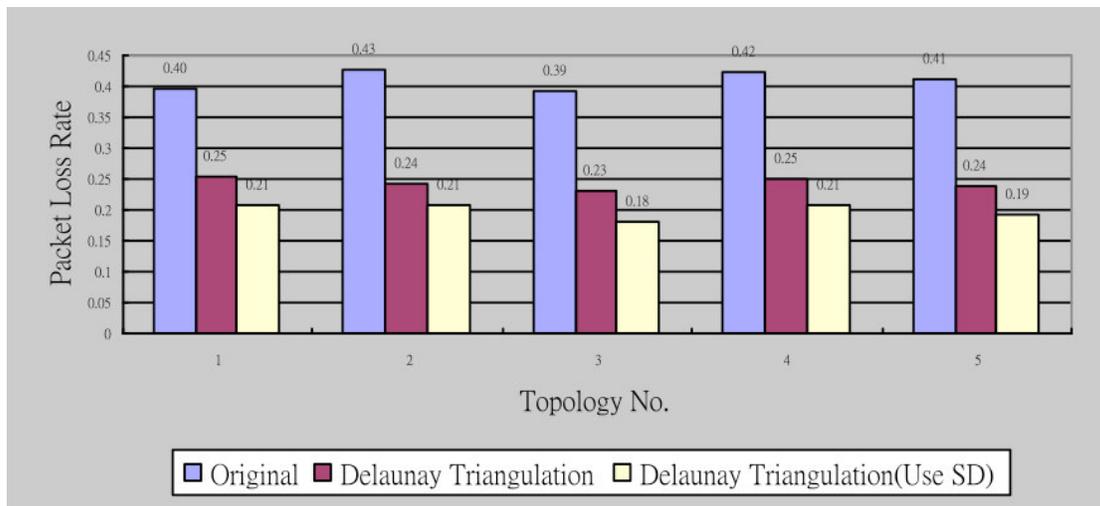

Figure 27. Comparison of packet loss rate

("200 nodes with 150 random incoming ftp connections").

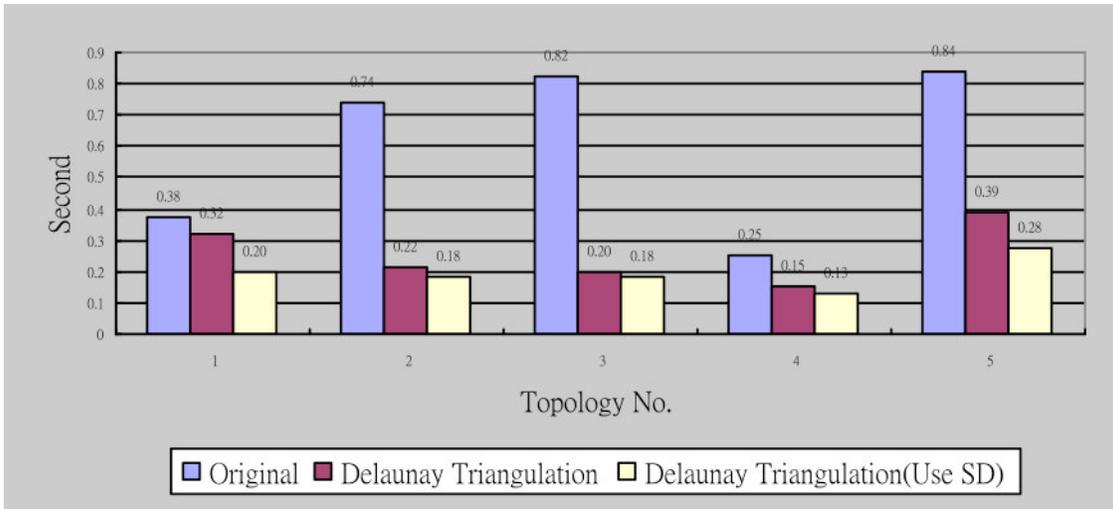

Figure 28. Comparison of delay time ("50 nodes with 30 random incoming ftp connections").

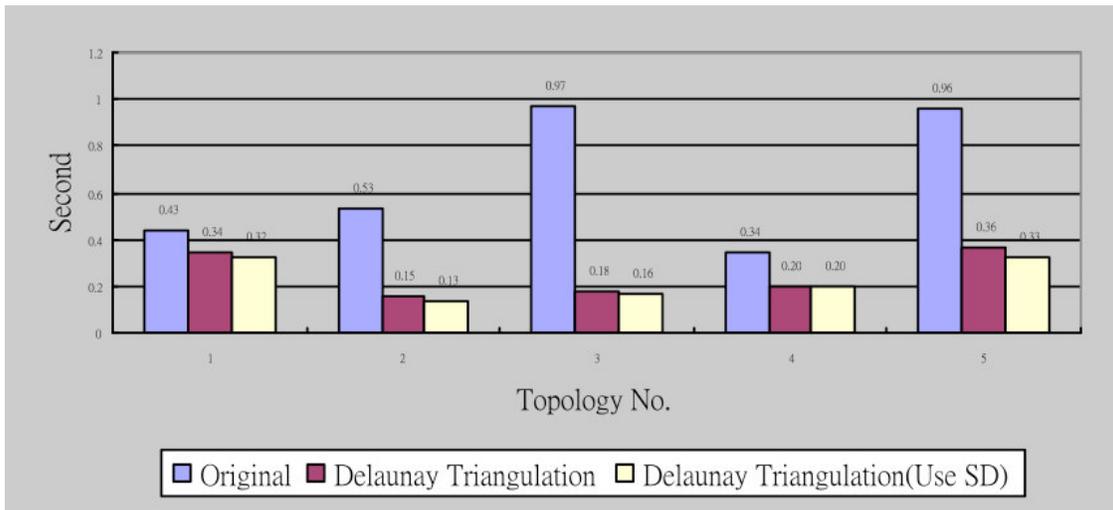

Figure 29: Comparison of delay time ("100 nodes with 50 random incoming ftp connections")

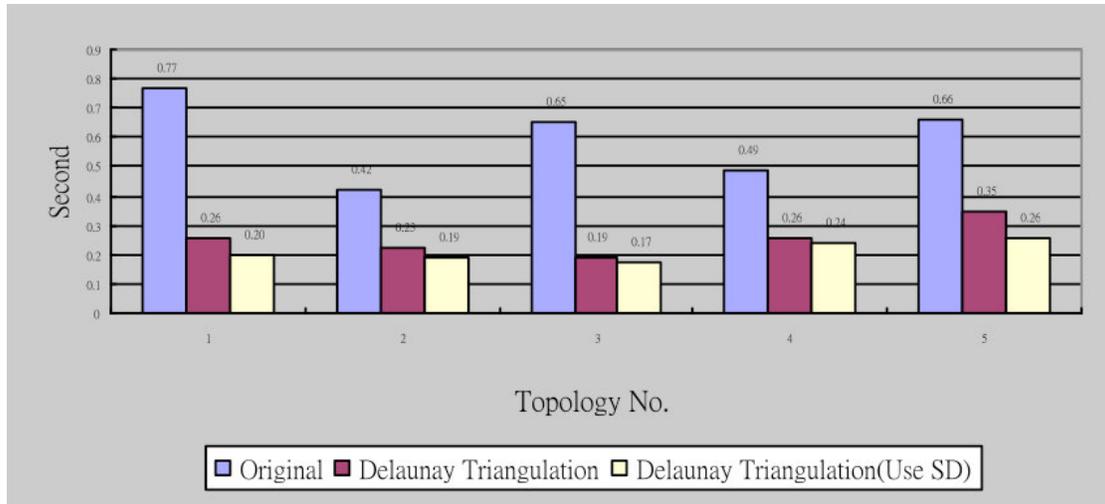

Figure 30. Comparison of delay time ("150 nodes with 100 random incoming ftp connections").

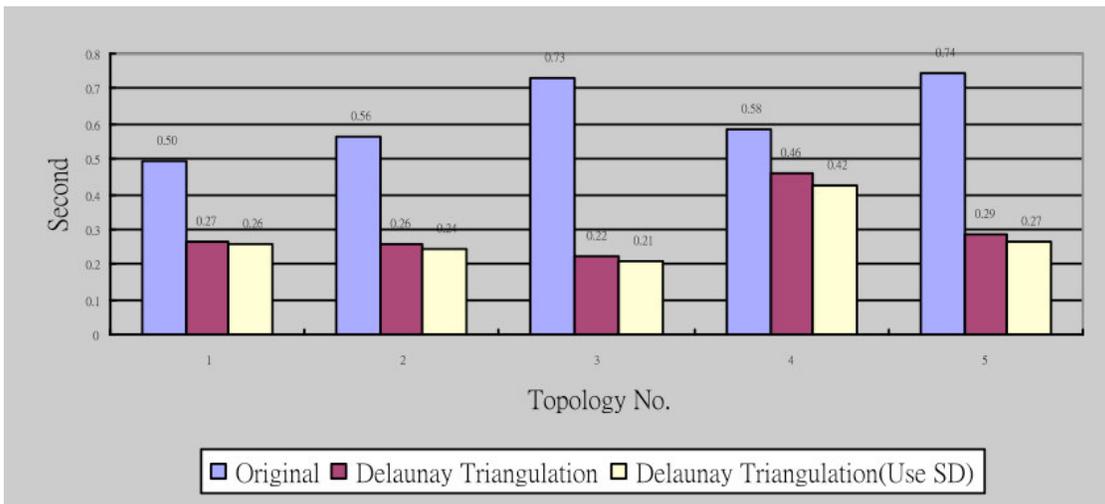

Figure 31. Comparison of delay time ("200 nodes with 150 random incoming ftp connections").

### 3.4 Performance Comparisons   Uniform Traffic

This category of simulations shows the comparisons of the "Original", "DT only", and "DT+SD" methods in terms of "number of nodes against throughput (bps)" (Figures 32-34). Figure 32 shows the comparison of number of nodes against throughput (bps). "DT+SD" can support more number of nodes than that of both "Original" and "DT only" at the same throughput. Figure 33 shows the comparison of packet loss rate against number of nodes. "DT+SD" has the least packet loss rate than that of both "Original" and "DT only" at the same number of nodes. Figure 34 shows the comparison of delay time against number of nodes. "DT+SD" has the least delay time than that of both "Original" and "DT only" at the same number of nodes.

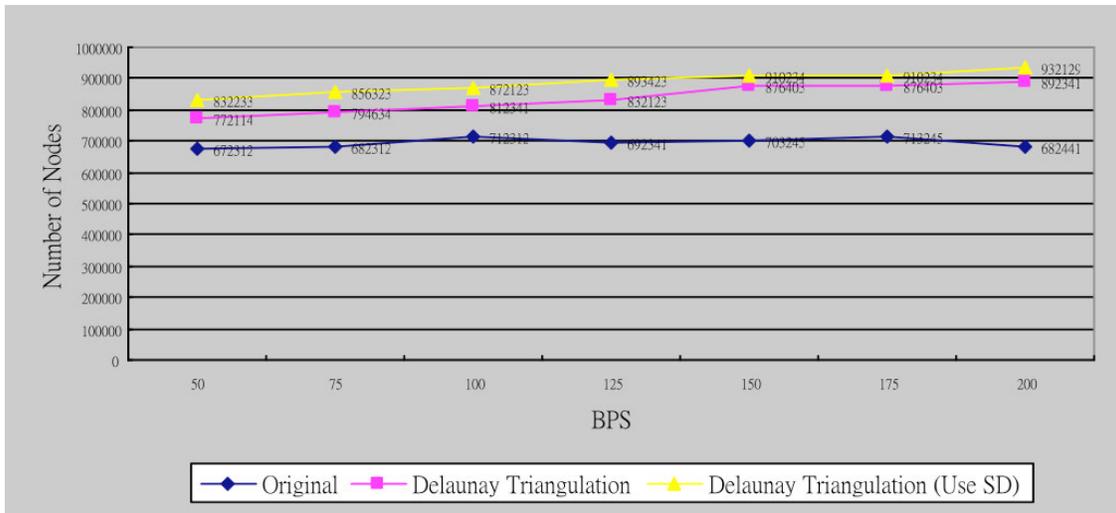

Figure 32. Comparison of number of nodes against throughput (bps).

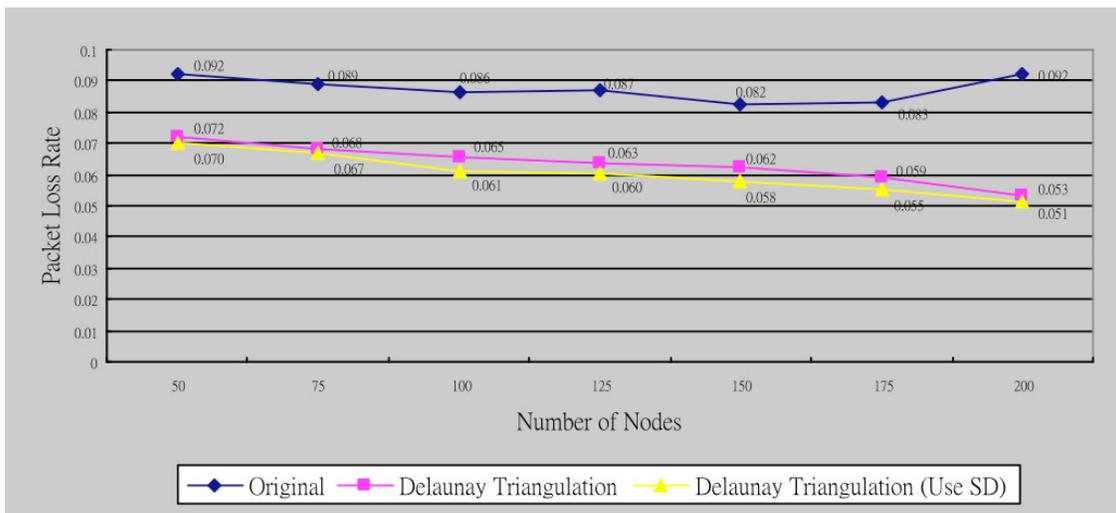

Figure 33. Comparison of packet loss rate against number of nodes.

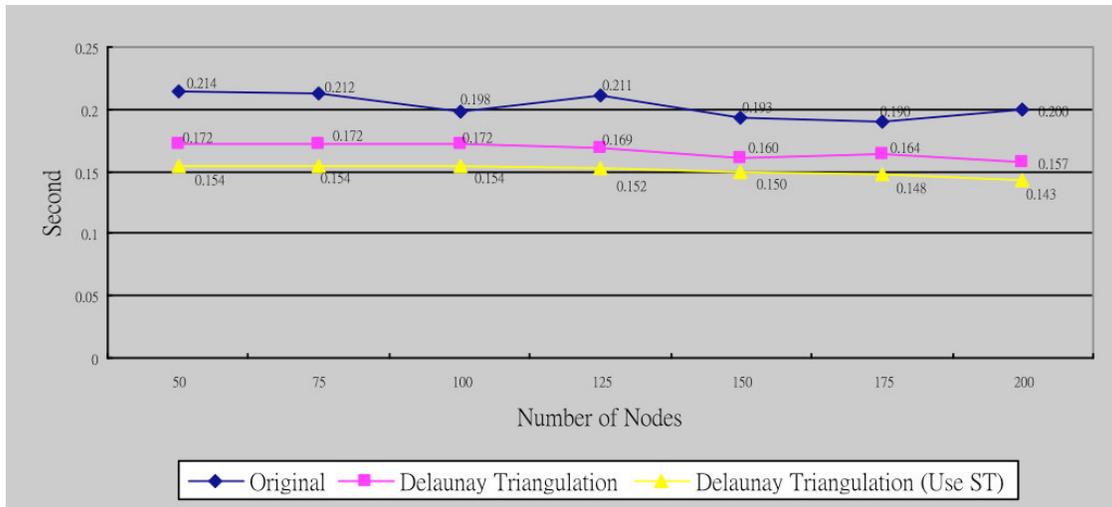

Figure 34. Comparison of delay time against number of nodes.

## 4. CONCLUSION

In this paper, we propose a framework using geometric algorithms to largely reduce the interference among nodes in a wireless mesh network. We first convert network problems into geometry problems in graph theory, and then solve the interference problem by geometric algorithms. We first define line intersection in a graph to reflect radio interference problem in a wireless mesh network. We then use plan sweep algorithm to find intersection lines, if any; employ Voronoi diagram algorithm to delimit the regions among nodes; use Delaunay Triangulation algorithm to reconstruct the graph in order to minimize the interference among nodes. Finally, we use standard deviation to prune off those longer links (higher interference links) to have a further enhancement. This hybrid solution is proved to be able to significantly reduce interference in $O(n \log n)$ time. Simulations show that the proposed framework is effective in increasing throughput, reducing both packets loss rate and delay time on average.

**Authors**

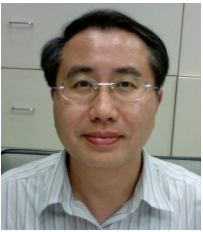

Hung-Chin Jang received his BS in Applied Mathematics from National Chengchi University, Taiwan, in 1984, MS in Mathematics, Statistics, and Computer Science, and Ph.D. in Electrical Engineering and Computer Science from University of Illinois at Chicago, U.S.A., in 1988 and 1992, respectively. He was an associate professor in Applied Mathematics, the Chair of Department of Computer Science, the Chair of Mater Program in Computer Science for Professional Education, National Chengchi University. Currently, he is an associate professor in Computer Science, the Director of Mobile Computing and Communication Lab., National Chengchi University. His current research interests include WLAN, Vehicular Ad Hoc Network (VANET), WiMAX, mobile communication systems, and mobile learning.